Modeling Bank Systemic Risk of Emerging Markets under Geopolitical Shocks: Empirical Evidence from BRICS Countries

Haibo Wang

Division of International Business and Technology Studies, A.R. Sánchez Jr. School of Business,

Texas A&M International University, Laredo, TX, USA, hwang@tamiu.edu

**Abstract**

In this study, we introduce an analytics framework, the Bank Risk Interlinkage with Dynamic Graph and Event Simulations (BRIDGES), to capture the systemic risks associated with the growing economic influence of the BRICS nations. This framework includes a Dynamic Time Warping (DTW) method to construct a dynamic network of 551 BRICS banks with their annual balance sheet data from 2008 to 2024; a trend analysis in risk ratios to detect shifts in banks' behavior; a Temporal Graph Neural Network (TGNN) to detect anomalous changes in the bank network's structural relationships; and Agent-Based Model (ABM) simulations to measure the impact of anomalous changes on network stability and assess the banking system's resilience to internal financial failure and external geopolitical shocks at the individual country level and across BRICS nations.

Our simulation results highlight several important insights. The failure of the largest BRICS banks can cause more systemic damage than that of financially vulnerable or anomalous banks due to the panic effects. Moreover, compared to the failure of the largest BRICS banks, a geopolitical shock with correlated country-wide propagation can cause more systemic damage, resulting in a near-total systemic collapse. Our findings suggest that the panic over the failure of the largest BRICS banks and large-scale geopolitical shocks are the primary threats to the financial stability of the BRICS nations, which traditional bank risk analysis models might not detect.

Keywords: Systemic Risk; Simulation; Emerging Markets; Agent-Based Modeling; Geopolitical Shocks; Financial Contagion

JEL Classification: C53; C55; F36; F52; G15; G21; G28; G32; O16

## 1. Introduction

The 2008 Global Financial Crisis (GFC) showed that global stability relies heavily on a small group of "too big to fail" institutions (Sorkin, 2010). The rise of the BRICS nations has significantly altered this landscape since then (Stuenkel, 2020). Today, BRICS banking systems house some of the world's largest financial entities by asset size (Armijo and Roberts, 2014). Domestic distress in the BRICS nations could trigger a large-scale emerging market crisis and create systemic risk through cross-border connections. Recent research points to the growing international footprint of BRICS banks across global trade finance and sovereign debt markets (Cerutti et al., 2018; Horn et al., 2021). For example, the five major Chinese lenders on the 2025 List of Global Systemically Important Banks (G-SIBs)[1] illustrate this rapid expansion. Bank of China (BoC) and Industrial and Commercial Bank of China (ICBC) have built massive global networks, reaching over 60 and 47 countries, respectively, while China Construction Bank (CCB), Agricultural Bank of China (ABC), and Bank of Communications (BoCom) focus their presence in 12 core global financial hubs (available in supplement materials). While BRICS nations are geographically disconnected, their banking systems are interconnected through two mechanisms. First, the BRICS nations provide an infrastructure fund for emerging markets through the New Development Bank (formerly the BRICS Development Bank[2]), which serves as a centralized hub. At the same time, BRICS nations deepened their financial ties through the BRICS Contingent Reserve Arrangement (CRA) and various currency swaps[3]. Therefore, severe financial distress in the BRICS nations can pose a threat of global contagion and disrupt global supply chains and capital flows of emerging markets.

Post-GFC risk measurement often relies on specialized analytical tools. For example, the SRISK framework estimates a bank's potential capital shortfall during future crises using high-frequency market data (Brownlees and Engle, 2017). Using market prices for risk measurement can provide forward-looking analysis if all banks were publicly listed, but it is impractical in this study, as many BRICS banks are privately owned or state-owned. In fact, even in the advanced economies, most banks are privately owned. For example, there are only 597 publicly listed US financial institutions out of 4415 on the FDIC list[4]. Most financial institutions in the BRICS nations are unlisted, state-owned entities, or have highly illiquid shares. It is not appropriate to use SRISK to calculate reliable market volatility for BRICS banks without high-frequency market data. Researchers have to work with annual balance sheet figures from commercial data providers on BRICS banks. The gap between available data and the forward-looking SRISK model limits its use for systemic risk analysis across BRICS banking systems.

To address the gap, the proposed framework, Bank Risk Interlinkage with Dynamic Graph and Event Simulations (BRIDGES), offers a practical solution. For each BRICS bank, we use a balance-sheet-based capital shortfall (CS) to calculate a baseline for systemic risk (SRISK_CS). To include a forward-looking analysis in BRIDGES, we develop a four-stage process. First, we choose the Dynamic Time Warping (DTW) distance metrics to measure the structural relationships between banks over time. Second, we implement a Temporal Graph Neural Network (TGNN) to detect anomalous shifts in the structures by evaluating long-term financial trajectories from the balance sheet. Third, we perform Agent-Based Model (ABM) simulations to test network stability in response to past extreme events and hypothetical future shocks. Fourth, we consolidate the results of previous stages into measurable impacts on systemic risk contagion. BRIDGES is designed to explore two research questions (RQs). RQ1 asks how the risk connections and network structures have changed over time within the BRICS banking systems. RQ2 examines the resilience of BRICS banking systems to geopolitical shocks and traces the paths of their risk propagation. The components in the four-stage process are implemented by the statistical and neural network libraries in Python.

---

[1] https://www.fsb.org/2025/11/2025-list-of-global-systemically-important-banks-g-sibs/, Accessed February 7th, 2026

[2] https://www.ndb.int/about-us/essence/history/, Accessed February 7th, 2026

[3] http://www.brics.utoronto.ca/docs/140715-treaty.html, Accessed February 7th, 2026

[4] https://banks.data.fdic.gov/bankfind-suite/bankfind, Accessed February 7th, 2026

The BRIDGES framework introduces a new empirical methodology for dynamic risk analysis that does not rely on high-frequency market data but uses annual balance sheet figures, enabling the testing of complex behavioral hypotheses. The simulation results show that banks with the highest SRISK_CS cause less systemic damage than banks with the largest assets under geopolitical shocks, suggesting that systemic collapse in BRICS might be driven by panic rather than individual banks' financial insolvency. The failure of very large banks (G-SIBs) acts as a public signal of the panic that triggers the systemic collapse. The simulation results also resolve the "Russian paradox," showing that Russian banks with low SRISK_CS, resulting from frequent state-led recapitalizations, are in fact associated with high network instability. The simulation of several geopolitical scenarios demonstrates that a correlated, country-wide geopolitical shock is more destructive than any internal financial insolvency.

The rest of the paper is organized as follows. Section 2 presents the theoretical background. Section 3 proposes the BRIDGES framework and describes the data source. Section 4 reports empirical results and discusses the main findings and implications. Finally, section 5 gives the conclusions.

## 2. Theoretical Background

### 2.1. Economic Theories

Economic models that account for emerging market volatility, state intervention, rapid credit expansion, and institutional constraints motivate the computational approach we adopt in this study. For example, BRICS nations experienced rapid credit growth after the GFC, which Hyman Minsky's Financial Instability Hypothesis can examine the risk cycles within these economies (Minsky, 1970). Minsky argued that extended periods of economic calm could encourage speculative lending. Within the BRICS nations, rapid credit expansion may be linked to debt-financed infrastructure projects and real estate booms. Orlik and Orlik (2020) found that the banking system in China may shift from stable "hedge" financing toward the "Ponzi" structures. Such vulnerability can be amplified by rapid credit growth (Bouvatier et al., 2012; Strobl, 2022). Thus, a domestic credit bust or an external event in the BRICS banking sectors, such as a commodity price crash affecting Brazil or Russia, can send shocks through both structural and behavioral channels. Additionally, large, state-owned banks (SOBs) in China and Russia dominate the market by raising loan rates (Lawson, 1994; Mengistae and Colin Xu, 2004; Ramamurti, 1987), leading to a modified "Too Big to Fail" problem (Stern and Feldman, 2004). These banks are encouraged to take risks without generating profit, but to fulfill government policy mandates, resulting in higher lending default rates. The moral hazard of these SOBs was examined by the Competition-Fragility hypothesis (Marcus, 1984) and the Competition-Stability hypothesis (Boyd and De NicolÓ, 2005). To examine these hypotheses, Standard econometric methods using high-frequency market data, such as Vector Autoregression (VAR) and Generalized Autoregressive Conditional Heteroskedasticity (GARCH) models, were used to model financial panics across structural and behavioral channels (An et al., 2022). However, these methods might struggle to examine the BRICS banking sectors due to a scarcity of market data, complex ownership structures of SOBs, and overlapping factors.

Thus, in this research, we develop an Agent-Based Model (ABM) to examine both channels with forward-looking analysis despite the lack of high-frequency market data for many BRICS banks. This approach creates a simulated laboratory for testing hypotheses. We chose this method because it effectively handles node heterogeneity and complex network feedback loops. BRICS banking sectors vary widely. The ABM allows us to model each bank as an individual agent based on its specific balance sheet data. If a single bank agent fails, the model can trace the direct domino effect on its counterparties. The simulation is likely to capture non-linear market reactions as well. For example, a distressed bank's fire sale of assets might depress market prices, thereby weakening the balance sheets of previously healthy banks.

We embedded the ABM within a Monte Carlo Simulation (MCS) framework to account for uncertainty. Rather than producing a single outcome, the MCS runs thousands of simulations. This generates a probability distribution of potential future states. Such an approach permits a probabilistic assessment of network resilience. By combining these methods, we use historical balance sheet data to conduct forward-looking stress tests. This integrated framework appears to offer a useful tool for policy analysis and financial planning.

### 2.2. Analytics Methods of BRICS Banks

Scholars evaluating the financial systems of BRICS nations generally select empirical methods adapted to the data limitations typical of emerging markets. A large portion of this literature relies on dynamic panel data techniques to model bank performance and economic growth. Revenue diversification, for instance, is likely to serve as a structural response to concentration risk. In their analysis, Moudud-Ul-Huq (2020) and Sharma and Anand (2018) suggested that BRICS banks may actually gain more from diversifying revenue streams than their ASEAN peers do. Umar and Sun (2016) and Umar et al. (2018) found evidence pointing to an inverse relationship between liquidity creation and funding liquidity. Their results imply that enforcing stricter capital requirements might inadvertently suppress liquidity creation. This observation appears to support the "financial fragility-crowding out" hypothesis.

To assess long-run relationships and cyclical co-movements, time-series approaches remain common. Using Autoregressive Distributed Lag (ARDL) and Vector Error Correction Models (VECM), some analysts found support for the supply-leading growth hypothesis. They also noted that global macroeconomic variables appear to be the primary triggers for sovereign risk (Sehrawat and Giri, 2015; Singh et al., 2021). Even so, these causal connections seem to vary across different contexts. Siva Kiran Guptha and Prabhakar Rao (2018) applied the Toda-Yamamoto test and observed no uniform causal link between finance and growth among the member states.

The GARCH-family models can detect strong asymmetries in volatility by analyzing market ties among BRICS banking systems and show that Brazil, Russia, India, and South Africa face an elevated risk of external liquidity shocks (An et al., 2022; Mensi et al., 2016). Further analysis to measure spillover indices with Delta Conditional Value at Risk (CoVaR) shows that Russia and South Africa are net transmitters of shocks, and non-traditional banking activities in the BRICS banking systems might raise overall systemic risk (Ahmad et al., 2018; Qin and Zhou, 2019; Zeb and Rashid, 2019). Along similar lines, Saliba et al. (2023) used quantile estimation to argue that political instability leads to worse non-performing loans.

Ahmad et al. (2021) and Li (2019) used index-based methods and Principal Component Analysis (PCA) to examine regulatory environments and found that BRICS regulators are likely tightening capital requirements to align with the Basel accords, while lowering entry barriers to encourage competition. However, this fast expansion might carry a potential downside. Through PCA, Barik and Pradhan (2021) argued that aggressive financial inclusion could actually harm financial stability. This negative effect is likely to be tied to an erosion of basic credit standards.

To address the financial fragility caused by the lack of high-frequency market data in emerging markets, researchers developed four accounting-based balance sheet models, including Z-score aggregation, accounting-based CoVaR, structural Merton models adapted for balance sheets, and related distance-to-default indicators (Huang et al., 2012; Lall, 2009). These models can detect vulnerabilities at the individual institution level but cannot capture time-varying spillovers at the system level. The proposed BRIDGES framework can build on this foundation by using accounting measures such as SRISK_CS as a baseline and leveraging TGNN and ABM simulation to advance from simple vulnerability identification to modeling systemic, time-varying contagion.

### 2.3. Information Complexity in Risk Analysis

Basic models often struggle to capture systemic risk and financial contagion. We design the BRIDGES framework around an escalating approach to information complexity, and this section outlines the theoretical rationale for its structure.

A zero-order model for balance sheet data provides a static snapshot. While such an approach might identify a bank with a high-risk score, it treats that institution as an isolated data point. It lacks a mechanism to gauge how a single failure is likely to impact the broader system. First-order models advance this by identifying linear trends from balance sheet data. They are incapable of detecting the sudden, accelerating crisis. The systemic events during the crisis, such as risk spillovers and behavioral panics, are driven by non-linear interactions within the system.

Thus, in financial applications, we need to design models to capture these non-linear interactions by mapping the channels of contagion to the systemic events of the crisis. Using nodes representing banks and edges denoting exposures or similarities, ABM can simulate how a shock within a few banks might propagate through the banking system under various crisis scenarios, thereby modeling behavioral patterns in crises, such as bank runs during the collapse of Silicon Valley Bank in 2023. In fact, ABM, as an approach that combines financial economics and computational methods, is often applied to analyze macroeconomic and monetary policy (Kukacka and Kristoufek, 2020; Lux, 2018; Vandin et al., 2022). In this study, we incorporate the ABM within the BRIDGES framework to assess BRICS banks' financial vulnerability and risk contagion.

### 2.4. Hypotheses Development

The central research question concerns the trigger of systemic collapse. If systemic risk is a function of structural contagion, then the failure of the most financially weak institutions should cause the most damage. The banks leading up to the GFC told a different story. Additionally, Large-scale geopolitical events, such as sanctions or trade wars, might trigger a breakdown of trust that does more damage than an individual bank's failure, highlighting the unique vulnerabilities of emerging economies to external shocks. Based on the theoretical difference between structural and behavioral contagion, two hypotheses are proposed for testing in the computational experiment. Our proposed BRIDGES framework can provide a controlled environment to test these two hypotheses.

**Hypothesis 1 (H1)**: The systemic damage caused by a bank's failure is more associated with its size than with its financial vulnerability.

**Hypothesis 2 (H2)**: The systemic impact of a large-scale country-wide correlated external shock is an order of magnitude greater than the impact of an internal failure of even the most systemically important banks.

## 3. Data and Methodology

### 3.1. Data Source and Variable Selection

In this study, we select 22 financial and 3 descriptive variables from BankFocus for BRICS commercial banks for the period 2008 to 2024. Our final sample of 551 banks is smaller than the thousands of microbanks operating in the BRICS nations and also smaller than the 734 registered banks with under 10% missing values across 25 variables in the database. This sample represents over 97.8% of total banking system assets in BRICS nations, and the country count is reported in Table 1. Excluding the long tail of micro-banks with poor reporting quality in this sample introduces a slight survivorship bias but fully captures the banks whose failure would lead to macroeconomic spillover. This trade-off helps focus on larger entities to maintain computational stability.

In this study, we applied strict data cleaning procedures to assemble a continuous time series required by DTW. Banks with more than 10% missing data over the five consecutive years have dropped. For the 551 banks, missing values were imputed using machine learning interpolation. Missing tail values were then estimated using cross-sectional imputation via a random forest algorithm within each institution's asset-size peer group. TGNN within the BRIDGE framework also partially mitigates this missing-value problem. TGNN updates a bank's representation by aggregating features from its network neighbors through message passing. As a result, the architecture is inherently reliable in the presence of isolated missing node features. This structure is likely to allow a bank's unobserved state to be inferred from its closest peers.

Table 1 details the variations across the BRICS banking systems. Such variation may suggest that relying on a single metric to evaluate their risk is likely inadequate across the BRICS nations. China's banking sector shows baseline stability and operates on a scale larger than that of other BRICS nations, while Russia's banking sector shows more volatility with more smaller banks. This instability can be observed in the mean Non-Performing Loans (NPL) ratio (0.446 for China and 4.830 for Russia), the NPL standard deviation (2.365 for China and 12.845 for Russia), and the CET1 ratios (20.909 for China and 301.258 for Russia). The standard deviation indicates year-to-year fluctuations and provides the baseline context for using the BRIDGES framework to examine the theoretical difference between structural and behavioral contagion.

**Table 1. Descriptive Statistics of BRICS Banking Systems (2008-2024)**

| Country | Number of Banks | Observations | Assets (B USD) | | NPL Ratio (%) | | CET1 Ratio (%) | |
|---|---|---|---|---|---|---|---|---|
| | | | Mean | Std Dev | Mean | Std Dev | Mean | Std Dev |
| Brazil | 88 | 789 | 39.915 | 103.792 | 1.216 | 4.120 | 10.586 | 22.291 |
| China | 224 | 2135 | 223.972 | 659.932 | 0.446 | 2.365 | 9.744 | 20.909 |
| India | 65 | 645 | 51.825 | 96.787 | 1.509 | 3.625 | 15.323 | 36.240 |
| Russia | 153 | 1148 | 16.781 | 62.973 | 4.830 | 12.845 | 14.487 | 301.258 |
| South Africa | 21 | 218 | 37.009 | 53.223 | 2.692 | 6.693 | 15.444 | 18.662 |

Table 2 presents the selected variables for the BRIDGES framework. Three risk metrics (CET1 ratio, NPL ratio, and SRISK_CS) are calculated to offer a comprehensive view of the health of BRICS banks at the country level. The descriptive results of the selected variables are provided in Table 3. Figure 1 highlights the importance of selected variables and provides an answer to **RQ1**, as the risk profiles of the BRICS nations are heterogeneous and can generate conflicting signals.

**Table 2. Variable Selection for BRIDGES Framework**

| Category | Variables |
|---|---|

| Capital Adequacy | Fitch Core Capital; Tier 1 Capital; Total Subordinated Debt on Balance Sheet; Core Tier 1 Regulatory Capital Ratio; Tier 1 Regulatory Capital Ratio; Total Regulatory Capital Ratio |
|---|---|
| Asset Quality | Impaired Loans; Gross Loans; Loan Loss Provisions |
| Profitability | Net Income; Net Interest Income; Net Int Rev / Avg Assets; Net Income/ Average Total Assets |
| Leverage | Average Assets; Total Equity; Total Liabilities; Total Customer Deposits; Total Assets |
| Risk | Impaired Loans(NPLs)/ Gross Loans; Risk Weighted Assets including floor/cap per Basel II; Risk Weighted Assets Credit Risk; Risk Weighted Assets Market Risk; Risk Weighted Assets Operational Market Risk |
| Description | NAME, Country Name; Ranking Year |

**Table 3. Descriptive Result of Variables in the BRIDGES Framework**

| **Variable** | **Median** | **Std Dev** | **Variable** | **Median** | **Std Dev** |
|---|---|---|---|---|---|
| Gross Loans | 9.0E+03 | 2.7E+05 | Total Subordinated Debt on Balance Sheet | 196.11 | 8483.2 |
| Total Assets | 1.7E+04 | 4.5E+05 | Risk Weighted Assets, including floor/cap per Basel II | 1.1E+04 | 2.7E+05 |
| Average Assets | 1.6E+04 | 4.3E+05 | Risk Weighted Assets Credit Risk | 607.11 | 2.3E+05 |
| Impaired Loans | 0 | 2373.549 | Risk Weighted Assets Market Risk | 10.09 | 4922.56 |
| Total Liabilities | 1.5E+04 | 4.1E+05 | Tier 1 Regulatory Capital Ratio | 12.04 | 173.46 |
| Total Equity | 1.3E+03 | 3.6E+04 | Total Regulatory Capital Ratio | 14.80 | 23.85 |
| Net Interest Income | 369.95 | 8558.19 | Net Int Rev / Avg Assets | 2.42 | 3.71 |
| Loan Loss Provisions | 31.97 | 2083.62 | Net Income/ Average Total Assets | 0.89 | 2.75 |
| Net Income | 110.57 | 4299.32 | Risk Weighted Assets Operational Market Risk | 37.83 | 1.8E+04 |
| Fitch Core Capital | 69 | 3.2E+04 | Core Tier 1 Regulatory Capital Ratio | 10.1 | 28.52 |
| Tier 1 Capital | 1.3E+03 | 3.5E+04 | Total Customer Deposits | 1.2E+04 | 3.6E+05 |

Note: The unit of the non-ratio variable is $1M.

Figure 1 shows the trends in three risk metrics and the volatility patterns of each country's banking sector. Brazil, China, India, and South Africa strengthened their capital buffers post-GFC, as measured by CET1 and NPL ratios. Russia's banking sector reported the most severe volatility in the NPL ratio, suggesting the impact of geopolitical shocks.

**Figure 1.** BRICS: Average CET1 and NPL Ratios Over Time by Country

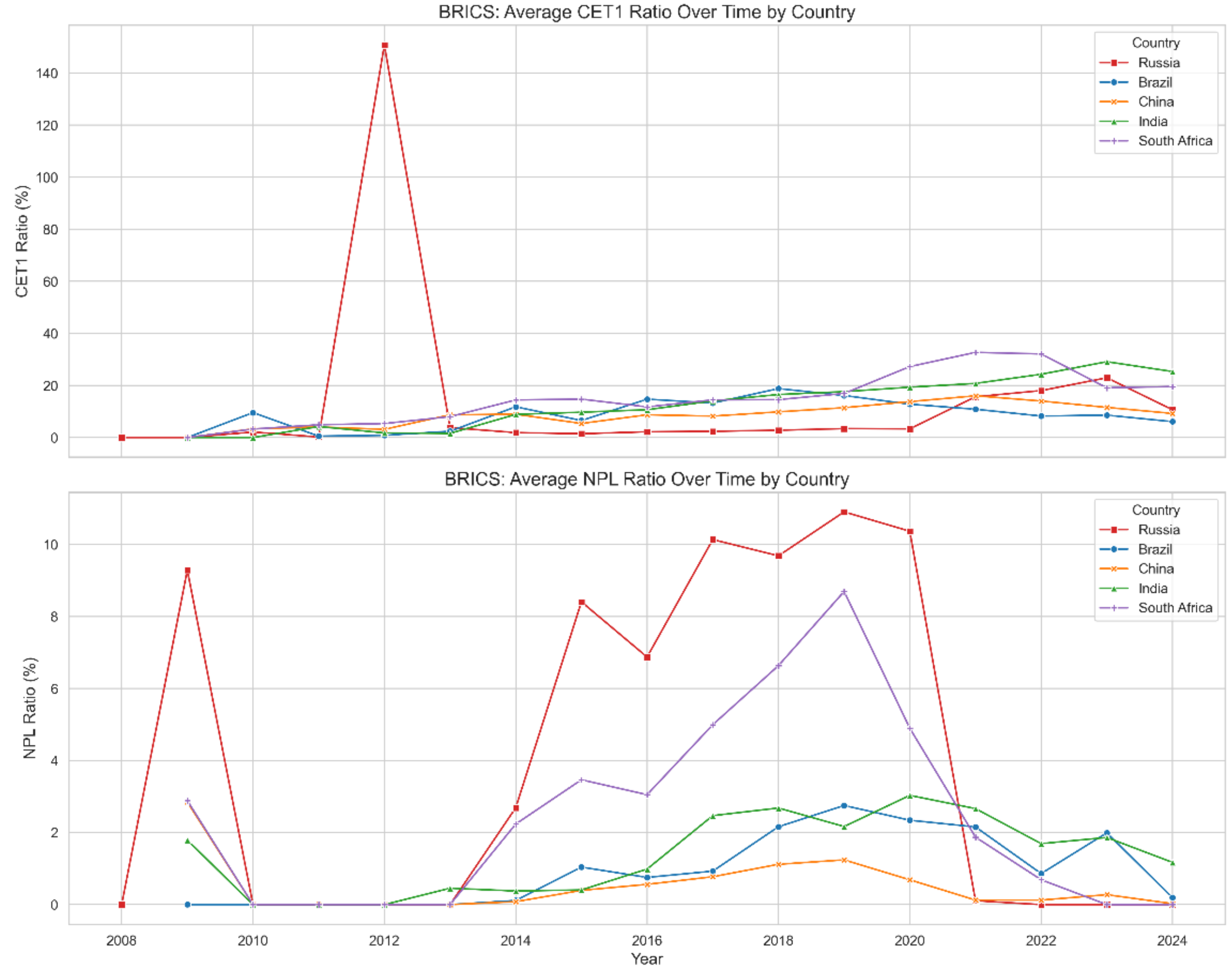

### 3.2. An Information-Theoretic Framework

The BRIDGES framework is structured to analyze risk based on the level of information complexity (zero-order, first-order, and second-order information). The workflow, illustrated in Figure 2, displays the information-theoretic approach through a three-stage pipeline. The process begins with the collection and preprocessing of annual balance sheet data (zero-order) and the calculation of key first-order risk measures for analysis in two streams: 1) DTW distances are calculated from balance sheet data to construct a sequence of dynamic networks where connections of nodes represent a bank's strategic similarity in stage one. To provide more insight into dynamic networks, the annual network is segmented by three bank asset sizes: mega-banks with assets of $50 billion or more, large banks with assets between $10 billion and $50 billion, and regular banks with assets of $10 billion or less. 2) Key risk measures are used to define a set of crisis scenarios for ABM simulations. Two streams merge within the core of the BRIDGES framework for analyzing second-order complexity. A TGNN was used to analyze how these network maps changed from one year to the next, learning the typical patterns of evolution to flag "anomalous" changes that could signal rising instability. TGNN identifies periods of anomalous structural change and the top anomalous banks within the BRICS banking systems in each of the two stages. With these dynamic triggers and two traditional risk measures, the BRIDGES framework moves from observation to experimentation. It utilizes ABM to examine how the system may respond to significant, real-world shocks in stage three. Using individual banks and depositors as agents, the ABM simulation of a virtual financial ecosystem is initiated by targeting specific banks as potential sources of instability under three scenarios: "too big to fail" based on the asset size, "most vulnerable" based on SRISK_CS, and top anomalous banks. ABM can perform comparative stress tests to assess the impact of shocks across three scenarios and model large-scale, correlated geopolitical events. Thus, the three stages of the BRIDGES framework can investigate second-order risk spillovers at both the bank and country levels, and validate the results through a reliability analysis.

**Figure 2**. BRIDGES Framework

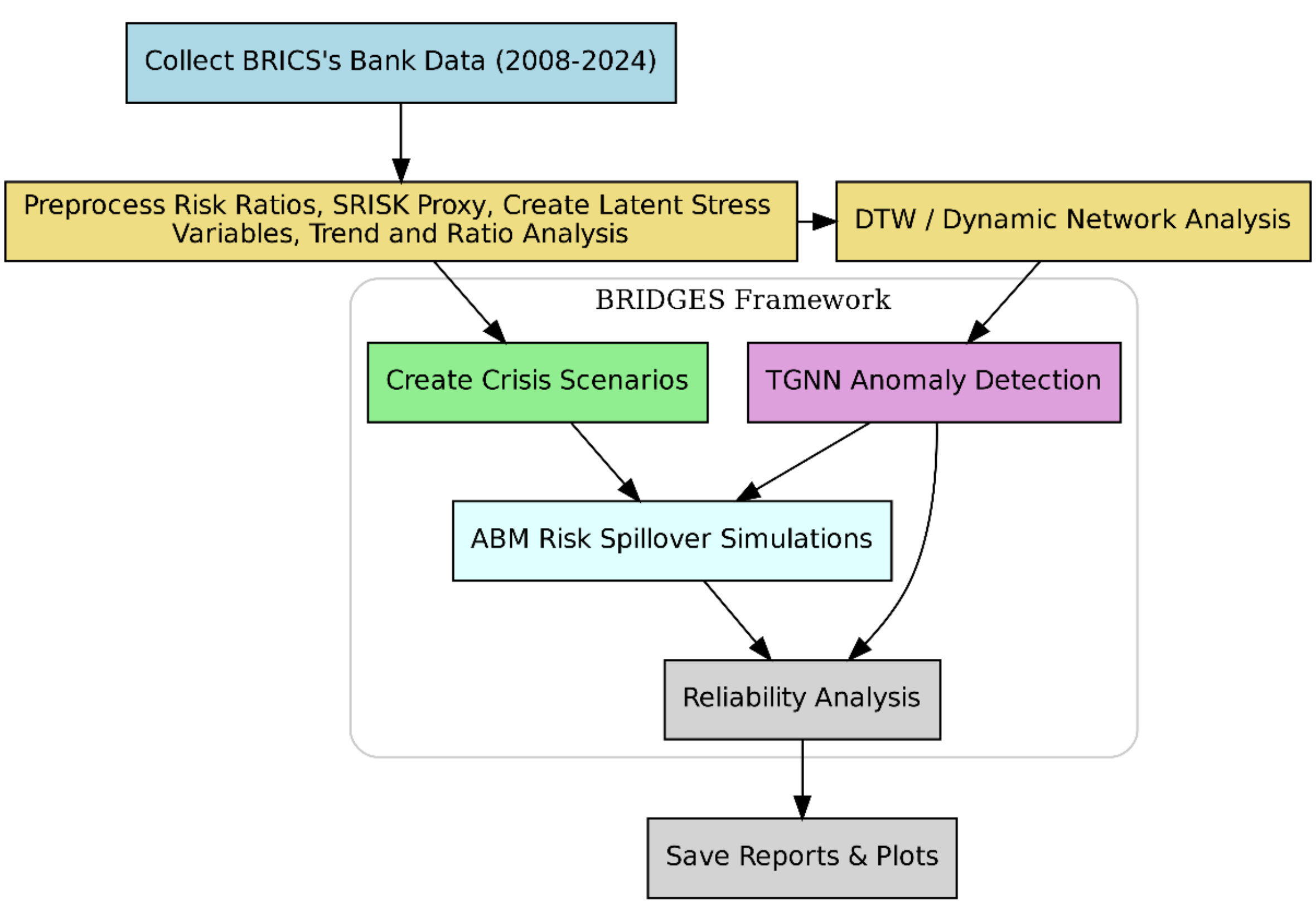

Source: Author's compilation.

### 3.3. Methodological Formulations

The detailed mathematical and computational formulations for the four components of the BRIDGES framework, SRISK_CS, DTW, TGNN, and ABM, are presented below.

#### 3.3.1. SRISK and SRISK_CS

The SRISK measure, developed by Acharya et al. (2012), Engle (2018), and Engle and Zazzara (2018), quantifies the capital shortfall a financial institution would likely face during a severe financial crisis.

The formal definition of SRISK for a firm *i* at time *t* is:

$$\text{SRISK}_{i,t} = k \cdot D_{i,t} - (1-k) \cdot W_{i,t} \cdot (1 - \text{LRMES}_{i,t}) \tag{1}$$

Where: $k$ is the prudential capital ratio, typically set at 8%. $D_{i,t}$ is the book value of the firm's total liabilities. $W_{i,t}$ is the market value of the firm's equity. $\text{LRMES}_{i,t}$ (Long-Run Marginal Expected Shortfall) is the source of the model's second-order property. It measures the expected percentage loss of the firm's equity conditional on a severe market decline.

The LRMES component captures the interactive dynamics between the firm and the market—its volatility and correlation—precisely when the system is under stress. It assesses how the bank is expected to behave within a failing system under dynamic market conditions. However, there is a conceptual flaw in equation 1, which allows a negative value of SRISK; FRB introduced SRISKv2[5] to address the flaw and use the maximum value between zero and SRISK to ensure no negative value of SRISK. However, the market data required by LRMES is unavailable for many banks in the BRICS nations. To address this constraint, this study employs balance-sheet-based SRISK_CS, which provides a broader focus on long-term structural risk rather than short-term market volatility.

The SRISK_CS used in the ABM simulations is a direct, accounting-based measure of a bank's capital shortfall relative to its current balance sheet position. The formula is:

$$\text{SRISK_CS}_{i,t} = k \cdot (\text{Total Liabilities}_{i,t}) - (1-k) \cdot (\text{Total Equity}_{i,t}) \tag{2}$$

Where: $k$ is the prudential capital ratio (set to 8%). $\text{Total Liabilities}_{i,t}$ and $\text{Total Equity}_{i,t}$ are the book values from the bank's annual balance sheet. Equation 2 compares required capital with available capital, which is the fundamental logic underlying SRISK, but it uses the balance sheet rather than high-frequency market data. SRISK_CS is a practical measure for detecting financially vulnerable banks and provides a baseline for generating inputs within the simulation framework for testing stress events.

[5] https://www.federalreserve.gov/econres/notes/feds-notes/sriskv2-a-note-20200918.html

SRISK in Equation 1 is based on the efficient-market hypothesis, assuming that equity prices and LRMES reflect a bank’s true vulnerability to systemic shocks. The use of SRISK_CS as a proxy for market-based SRISK represents a paradigm shift by theoretically stripping away the market response to shocks. Using TGNN and ABM in the BRIDGES framework does not reflect the market response to shocks; it shifts the paradigm from a market-based risk measure to a model-driven simulation of crisis mechanics. Given that the largest banks in China, Russia, India, and Brazil are state-owned, and state intervention distorts market signals in response to shocks, SRISK_CS provides a practical risk measure with a theoretical foundation for crisis. Therefore, the BRIDGES framework answered the question, *'Given this bank's structural weakness, how will its failure mechanically and behaviorally propagate through the system?* instead of *‘'What does the market expect will happen to this bank in a crisis?”*

To compensate for missing the market response to shocks in SRISK_CS, the BRIDGES framework embeds the forward-looking analysis to enhance its performance on detecting systemic risks in several ways: 1) Early warning with TGNN: While the SRISK_CS only represents a bank's structural vulnerability at a given year, TGNN can provide the dynamic context of structural vulnerability over time by identifying years of anomalous change in the network structure and report the year when a bank’s high SRISK_CS becoming systemically relevant. 2) Crisis Conditioning via ABM: SRISK uses the LRMES to model how a bank's equity will behave in a crisis. ABM achieves the same goal by examining behavior through direct simulation. 3) Scenario-Specific Stress Testing: ABM enhances the performance of detecting systemic risk by moving beyond a generic "market crash" to model-specific, tailored shocks. The impact of a bank's SRISK_CS may differ between a geopolitical sanctions scenario and a global systemic shock scenario. 4) Interrogating the SRISK_CS's Validity: The BRIDGES framework does not accept the SRISK_CS by default as the best measure of risk. Instead, the framework assesses its predictive power by running parallel simulations, including "too big to fail” scenarios (by Assets). The simulation results consistently show that shocking the largest banks often causes more systemic damage than shocking those with the highest SRISK_CS. This comparative analysis is a powerful enhancement for the limitations of the SRISK_CS. It demonstrates that in many crises, the panic effects triggered by the failure of a very large bank (G-SIBs) are more destructive than those triggered by the failure of a merely undercapitalized one.

### 3.3.2. DTW-Based Dynamic Network Analysis

In this study, we construct the evolving network of the BRICS banking systems using annual balance sheet data by using DTW rather than traditional methods like Euclidean distance and cosine similarity, which are common in other research (Chu et al., 2020; Girardi et al., 2021; Richards et al., 2008). For a given year *t*, we define a historical window of *w* years (*w*=5 in this study). For each bank $i$, we create a multivariate time series $S_{i,t}$ of its $K$ latent variables over this window: $S_{i,t} = (L_{i,t-W+1}, L_{i,t-W+2}, \ldots, L_{it})$, where $K$ is the number of latent variables tracked for each bank in each year. $L_{it}$ is K-dimensional latent-variable vector for bank $i$ at time $\tau$.

We compute the pairwise multivariate DTW distance between every pair of banks (*i,j*). DTW, a well-established technique for measuring similarity between time series (Berndt and Clifford, 1994), finds the optimal alignment by constructing a $W \times W$ cumulative distance matrix $D$, where an element $D(s, w)$ is given by the recurrence relation:

$$\mathrm{D(s,w) = d(L_{is}, L_{jw}) + min\{D(s-1,w), D(s,w-1), D(s-1,w-1)\}} \quad (3)$$

where $d(L_{is}, L_{jw})$ is the Pointwise distance between the latent-state vectors of banks $i$ and $j$ at aligned window positions $s$ and $w$. $(s, w)$ is the within-window time-step indices used by the DTW alignment algorithm, where $s, w \in \{1, \dots, W\}$. $D(s, w)$ is the minimum cumulative alignment cost up to the time-step pair $(s, w)$. The final distance, $DTW(S_{i,t}, S_{j,t})$, is the DTW distance between the two multivariate sequences for banks $i$ and $j$ in year $t$, equal to the terminal cumulative cost $D(W, W)$.. This distance is converted into a similarity score:

$$\mathrm{Sim_t(i,j) = exp(-\gamma \cdot DTW(S_{i,t}, S_{j,t}))} \quad (4)$$

where $\gamma$ is a scaling parameter. This results in a time-varying, weighted adjacency matrix $A_t$ where $A_t(i,j) = Sim_t(i,j)$, from which network centrality measures can be calculated for each year $t$. Since banks do not always report on the same schedule and often miss quarters or alternate reporting cycles, a strict point-for-point comparison would be invalid when we used Euclidean distance or cosine similarity to measure network edges. The DTW distance generally performs better on financial report data than other measures, as it warps the time axis to find the best-fit alignments of varying lengths, which capture strategic similarities rather than temporal alignment. In fact, bank financial activities and events, such as mergers or delayed responses to regulatory changes, might lead to poor performance for Euclidean distance and cosine similarity due to distortions or time lags introduced by these activities and events. DTW is less sensitive to inherent noise in financial data, including minor reporting discrepancies and temporary market fluctuations, and can recognize similar patterns even when they are uncoordinated.

The DTW network maps similarities in balance sheet trajectories by focusing on the overall shape of a bank's financial trend rather than individual data points, making it more likely that the network reflects genuine economic relationships rather than data artifacts. These trajectories economically represent common asset exposure and shared business models. Researchers studied information contagion and fire-sale spillovers and found that banks with highly correlated balance sheet evolutions are likely to be affected by the same macroeconomic factors and asset classes (Allen et al., 2012; Cai et al., 2018). If one bank suffers a shock that forces asset liquidation, other banks in the peer group with high DTW similarity are likely to experience simultaneous mark-to-market losses and correlated depositor runs. This makes the DTW network a highly relevant topology for simulating behavioral contagion.

### 3.3.3. TGNN for Anomaly Detection

With its dynamic network structure, BRIDGES utilizes second-order information by adopting a TGNN based on the Graph Convolutional Network (GCN) (Pareja et al., 2020), which updates its representation by aggregating information from the node's neighbors. Therefore, a bank's status can be contextualized by the concurrent states of its peers. The standard GCN layer is expressed as:

$$\mathrm{H}_t^{(\ell+1)} = \sigma(\widetilde{\mathrm{D}}_t^{-1/2}\widetilde{\mathrm{A}}_t\widetilde{\mathrm{D}}_t^{-1/2}\mathrm{H}_t^{(\ell)}\mathrm{W}^{(\ell)}) \tag{5}$$

, where $\ell$ is the graph convolution layer index. $\mathrm{H}_t^{(\ell+1)}$ is the updated node embedding matrix at the time $t$ after one GCN layer. $\sigma$ is an elementwise nonlinear activation function is applied after graph aggregation. $\widetilde{\mathrm{A}}_t$ is an adjacency matrix with self-loops added. $\mathrm{W}^{(\ell)}$ is the Trainable weight matrix for the graph convolution layer $\ell$. $\widetilde{\mathrm{D}}_t$ is a diagonal degree matrix associated with $\widetilde{\mathrm{A}}_t$, where $\widetilde{\mathrm{D}}_t(i,i) = \sum_j \widetilde{\mathrm{A}}_t\,(i,j)$. $\widetilde{\mathrm{D}}_t^{-1/2}\widetilde{\mathrm{A}}_t\widetilde{\mathrm{D}}_t^{-1/2}$ is the symmetrically normalized adjacency operator used by the GCN to aggregate neighborhood information.

To account for the temporal nature of the data, the EvolveGCN architecture learns how these second-order dependencies change over time by updating the weight matrix $W^{(l)}$ at each time step $t$ using a Recurrent Neural Network (RNN):

$$W_t^{(l)} = \mathrm{GRU}(W_{t-1}^{(l)}, \text{learning_signal}) \tag{6}$$

The final representations for each bank in the GCN show the level of isolation among nodes (banks), which is encoded as an anomaly score and may indicate anomalous behavior among peers.

While raw data is tabular, utilizing graph architecture is essential because systemic risk is fundamentally a network phenomenon. The tabular data creates the DTW distance matrix, which acts as a map of latent contagion channels. The TGNN processes this graph to understand how a shock to one row of tabular data (a bank) influences another through these hidden channels. The EvolveGCN model was trained in over 100 epochs using the Adam optimizer with a learning rate of 0.001. The architecture used two GCN layers with a hidden dimension of 64, and temporal evolution was handled by a Gated Recurrent Unit (GRU) cell. In this study, we applied an 80-20 train-test split. This setup is likely to minimize data leakage and yield reliable out-of-sample temporal predictions.

### 3.3.4. Agent-Based Model for Simulations

In this study, we structure the agent-based model (ABM) in BRIDGES using the Diamond and Dybvig (1983)'s theory and Bookstaber (2017)'s computational approach. The ABM simulates depositor-driven contagion by generating $N$ BankAgent objects to represent the banking networks. To reflect realistic conditions, each agent is initialized with balance sheet data from a specific year to ensure the simulation is in historical financial states. At the start of the simulation ($\tau = 0$ ), a shock is applied to a targeted set of banks $S$, reducing their capital by a percentage shock_pct:

$$C_i(0) \to C_i(0) \cdot (1 - \text{shock_pct}) \text{ for } i \in S \quad (7)$$

, where $C_i(0)$ is the initial capital of the bank $i$ at the start of the simulation.

The contagion mechanic is a behavioral fear factor, $F_i(\tau)$, for each bank $i$. $F_i(\tau)$ increases with capital loss but does not subside:

$$F_i(\tau) = \max\left(F_i(\tau - 1), 1 - C_i(\tau)/C_i(0)\right) \quad (8)$$

, where $C_i(\tau)$ is the capital of the bank $i$ at simulation time $\tau$.

Deposit withdrawals for each bank $i$ are then driven by a combination of its own fear and the systemic fear across all banks, which is $F_{sys}(\tau) = (1/N) \sum_j F_j(\tau)$. $F_{sys}(\tau)$ is the system-wide average fear level at the time $\tau$.

The Deposit withdrawals are:

$$W_i(\tau) = D_i(\tau - 1) \cdot [\alpha \cdot F_i(\tau - 1) + (1 - \alpha) \cdot F_{sys}(\tau - 1)] \cdot \psi \quad (9)$$

, where $W_i(\tau)$ is the withdrawal amount experienced by the bank $i$ at time $\tau$, $D_i(\tau - 1)$ is the deposit base of the bank $i$ carried into the time step $\tau$. In this equation, $\alpha$ weights the relative importance of individual versus systemic fear, while $\psi$ acts as a general panic sensitivity parameter. These withdrawals first deplete a bank's cash reserves. When those run out, the bank must sell its assets at a discount. This incurs losses and further erodes its capital. We consider the bank $i$ insolvent and remove it from the simulation if its capital, $C_i(\tau)$ falls to or below zero.

The behavioral parameters for fear ($\alpha$, $\psi$) rely on the global bank run literature. This prior research models how public signals appear to coordinate depositor actions (Goldstein and Pauzner, 2005). Historical balance sheet data cannot perfectly calibrate these variables. Relying on a single static assumption would likely be ad hoc. To resolve this, we embedded the ABM within a Monte Carlo simulation using the BRIDGES framework. In this study, we ran thousands of simulations across a wide, continuous distribution of panic sensitivities ($\psi$) and idiosyncratic-versus-systemic weightings ($\alpha$). The resulting expectations span a broad range of parameter settings to ensure that our outcomes reflect the structural features of the banking network rather than the result of isolated modeling choices.

## 4. Results and Discussion

### 4.1. Financial Interconnectedness and Risk

We group the BRICS bank networks into mega-, large-, and regular banks by asset size in Figure 3. Chinese banks dominate the mega bank network, which aligns with the list of Global Systemically Important Banks (G-SIBs) published by the Financial Stability Board (FSB, 2024).

Indian and Brazilian banks exhibit a high concentration within their large bank networks. In contrast, the regular bank network contains a mix of banks from across the BRICS nations.

**Figure 3**. BRICS Banks DTW Network Analysis by Asset Category

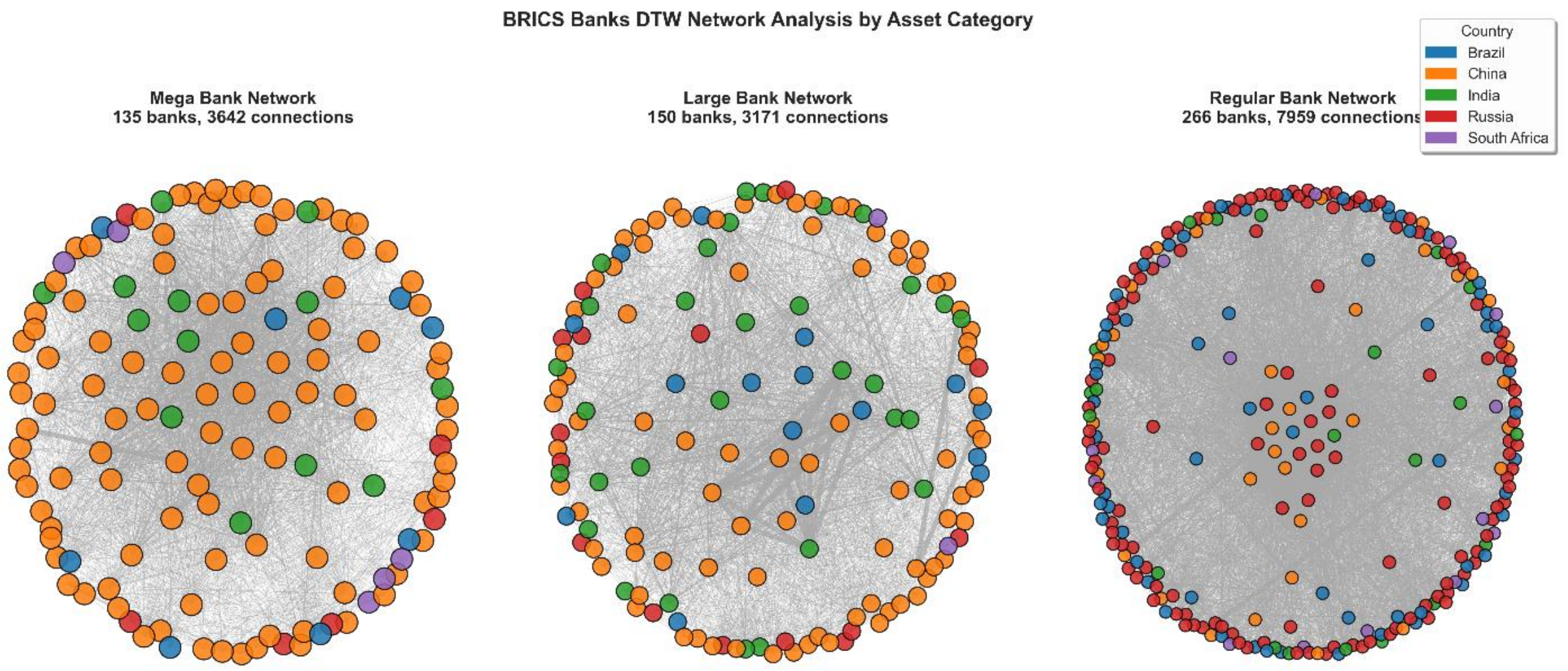

In the top panel of Figure 4, we observe systemic interconnectedness measured by the average rolling correlation of bank profitability. During the Post-GFC period, correlations increased and became more volatile. In the bottom panel, China's aggregated SRISK_CS exceeds the combined total of all other BRICS nations. Around 2014, Russia exhibited highly volatile interconnectedness while maintaining the lowest SRISK_CS throughout the period, often remaining near zero. This paradox is likely linked to frequent state-led recapitalizations implemented in response to international sanctions, suggesting that the Russian banks' primary risk depends heavily on the country's fiscal capacity. As a result, SRISK_CS might fail to capture this underlying vulnerability.

### 4.2. Resilience and Dynamics

In this study, we run simulations to examine how contagion spreads from internal financial failures. We select a small group of banks based on three distinct criteria. First, we choose the five largest banks by assets to represent the "too big to fail" category. Second, we select the five most vulnerable banks according to SRISK_CS. Third, we include banks flagged as highly anomalous by the baseline model and TGNN in Tables 4 and 5. The baseline model often detects smaller banks with short-term distress, while TGNN identifies banks showing unusual long-term patterns. We designed our ABM simulations to address the questions raised by this discrepancy between the two models. Systemic crises might begin with highly visible cases of distress. Alternatively, they are likely to emerge from sudden defaults by unpredictable banks.

Figure 4. BRICS: Evolution of Systemic Interconnectedness and Aggregated Systemic Risk

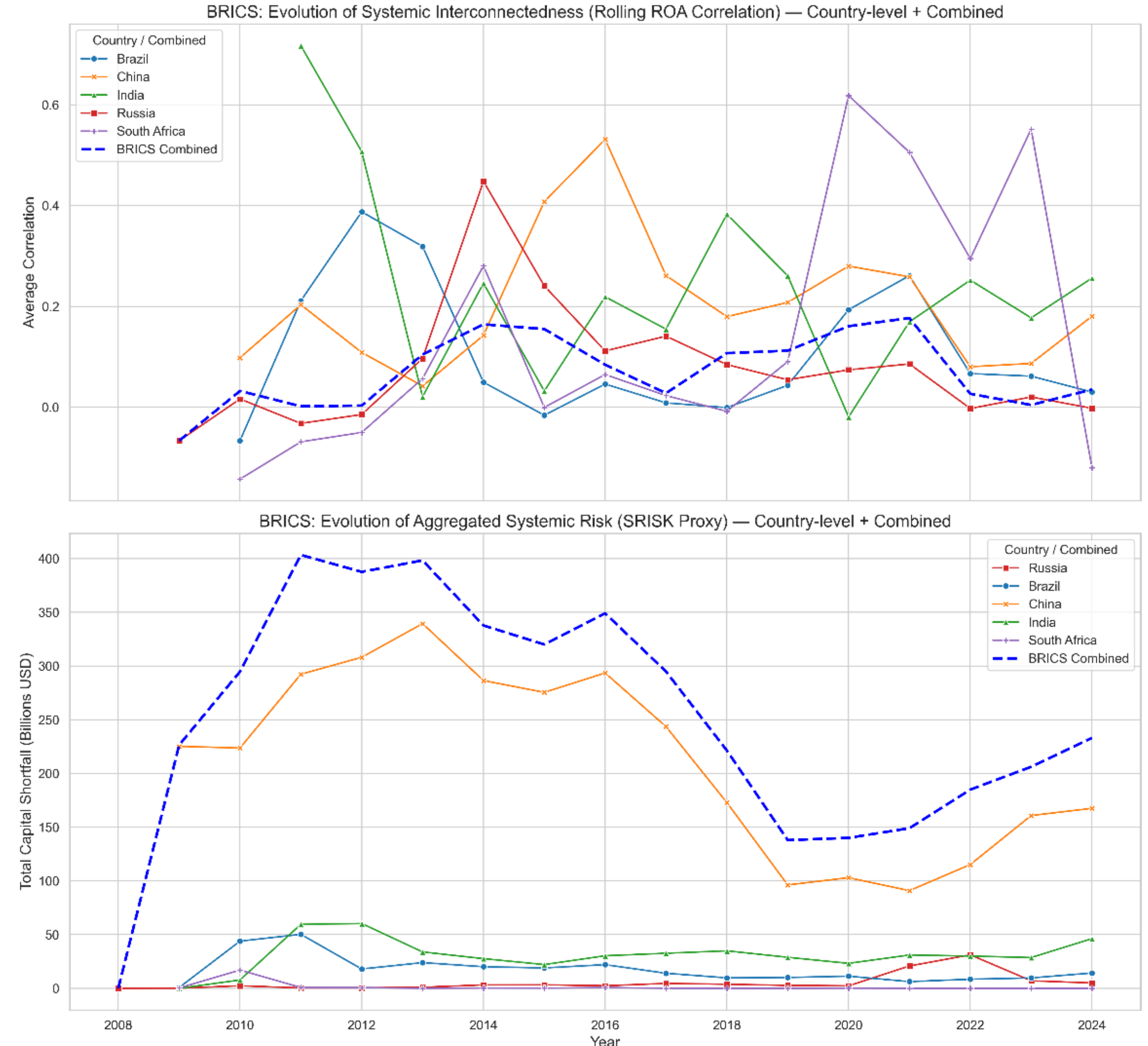

**Table 4. Comparison of Systemically Important Banks by First-Order Metrics**

| | Too Big To Fail (Top Assets) | | SRISK_CS | |
|---|---|---|---|---|
| | Name | Country | Name | Country |
| 2014 | Industrial and Commercial Bank of China | China | Postal Savings Bank of China | China |
| | China Construction Bank Corporation | China | Agricultural Bank of China | China |
| | Agricultural Bank of China | China | Industrial and Commercial Bank of China | China |
| | Bank of China | China | China Construction Bank Corporation | China |
| | Postal Savings Bank of China | China | Industrial Bank | China |
| 2018 | Industrial and Commercial Bank of China | China | Postal Savings Bank of China | China |
| | China Construction Bank Corporation | China | Agricultural Bank of China | China |
| | Agricultural Bank of China | China | State Bank of India | India |
| | Bank of China | China | Industrial Bank | China |
| | Bank of Communications | China | Bank of Communications | China |
| 2020 | Industrial and Commercial Bank of China | China | Postal Savings Bank of China | China |
| | China Construction Bank Corporation | China | State Bank of India | India |
| | Agricultural Bank of China | China | China Citic Bank | China |
| | Bank of China | China | China Zheshang Bank | China |
| | Postal Savings Bank of China | China | Caixa Economica Federal | Brazil |
| 2022 | Industrial and Commercial Bank of China | China | Postal Savings Bank of China | China |
| | China Construction Bank Corporation | China | Bank Trust | Russia |
| | Agricultural Bank of China | China | State Bank of India | India |
| | Bank of China | China | China Zheshang Bank | China |
| | Postal Savings Bank of China | China | VTB Bank | Russia |

**Table 5. Comparison of Anomalous Banks by Static vs. Dynamic Models**

| | Baseline Model | | TGNN | |
|---|---|---|---|---|
| | Name | Country | Name | Country |
| 2014 | State Street Corporation Brasil | Brazil | OCBC Wing Hang Bank | China |
| | International Bank of Azerbaijan Moscow | Russia | Macau Chinese Bank | China |
| | Finprombank | Russia | Locko Bank | Russia |
| | SMP Bank | Russia | Barclays Bank Brazil | Brazil |
| | Ubank | South Africa | NatWest Markets India | India |
| 2018 | India Post Payments Bank | India | OCBC Wing Hang Bank | China |
| | NatWest Markets India | India | Banco Inbursa Brasil | Brazil |
| | Banco Inbursa Brasil | Brazil | Tai Yau Bank | China |
| | International Bank of Azerbaijan Moscow | Russia | Tai Sang Bank | China |
| | Commercial Bank for Charity and Spiritual Development of Fatherland Peresvet | Russia | India Post Payments Bank | India |
| 2020 | International Bank of Azerbaijan Moscow | Russia | OCBC Wing Hang Bank | China |
| | Tai Yau Bank | China | Livi Bank | China |
| | NatWest Markets India | India | NatWest Markets India | India |
| | CentroCredit Bank | Russia | Goldman Sachs Asia Bank | China |
| | Investment Trade Bank | Russia | Banco KDB Brasil | Brazil |
| 2022 | Bank Trust | Russia | OCBC Wing Hang Bank | China |
| | NatWest Markets India | India | NatWest Markets India | India |
| | Bank Tavricheskiy | Russia | FirstRand India | India |
| | Tyme Bank | South Africa | Tyme Bank | South Africa |
| | FirstRand India | India | Tai Yau Bank | China |

Figure 5 presents the analytics results for 2014, 2018, 2020, and 2022, which suggest that failures among the top five banks by assets consistently involve the largest Chinese state-owned banks (G-SIBs), whose collapses trigger the most severe systemic damage across the BRICS banking systems. This outcome may suggest that the "too big to fail" problem is not strictly a financial issue or a structural flaw. Rather, it is a behavioral phenomenon. When G-SIBs fail, they send a strong second-order distress signal to the entire BRICS banking systems. This event breaks depositor confidence and sparks widespread panic.

**Figures 5. ABM Spillover Scenarios for 2014, 2018, 2020, and 2022**

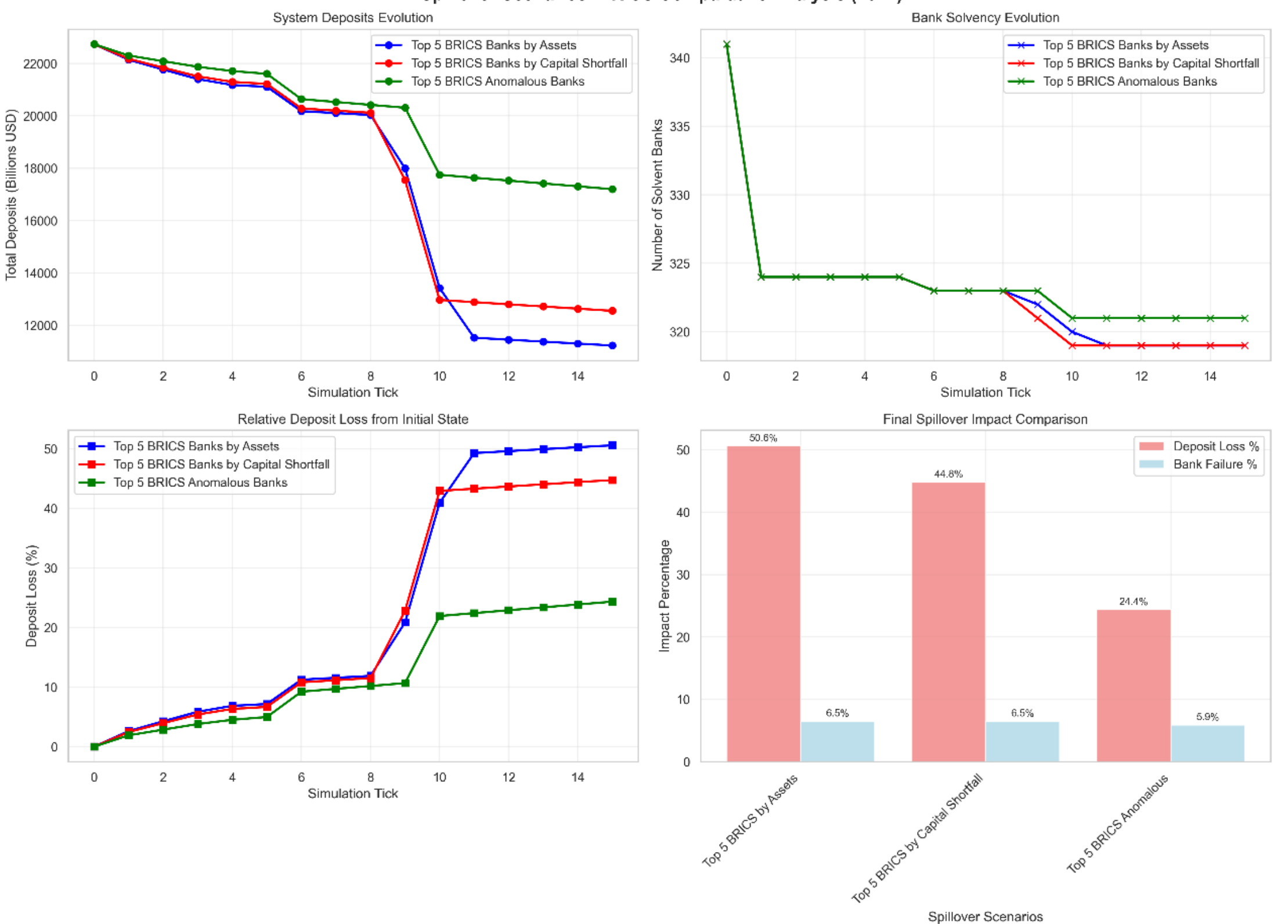

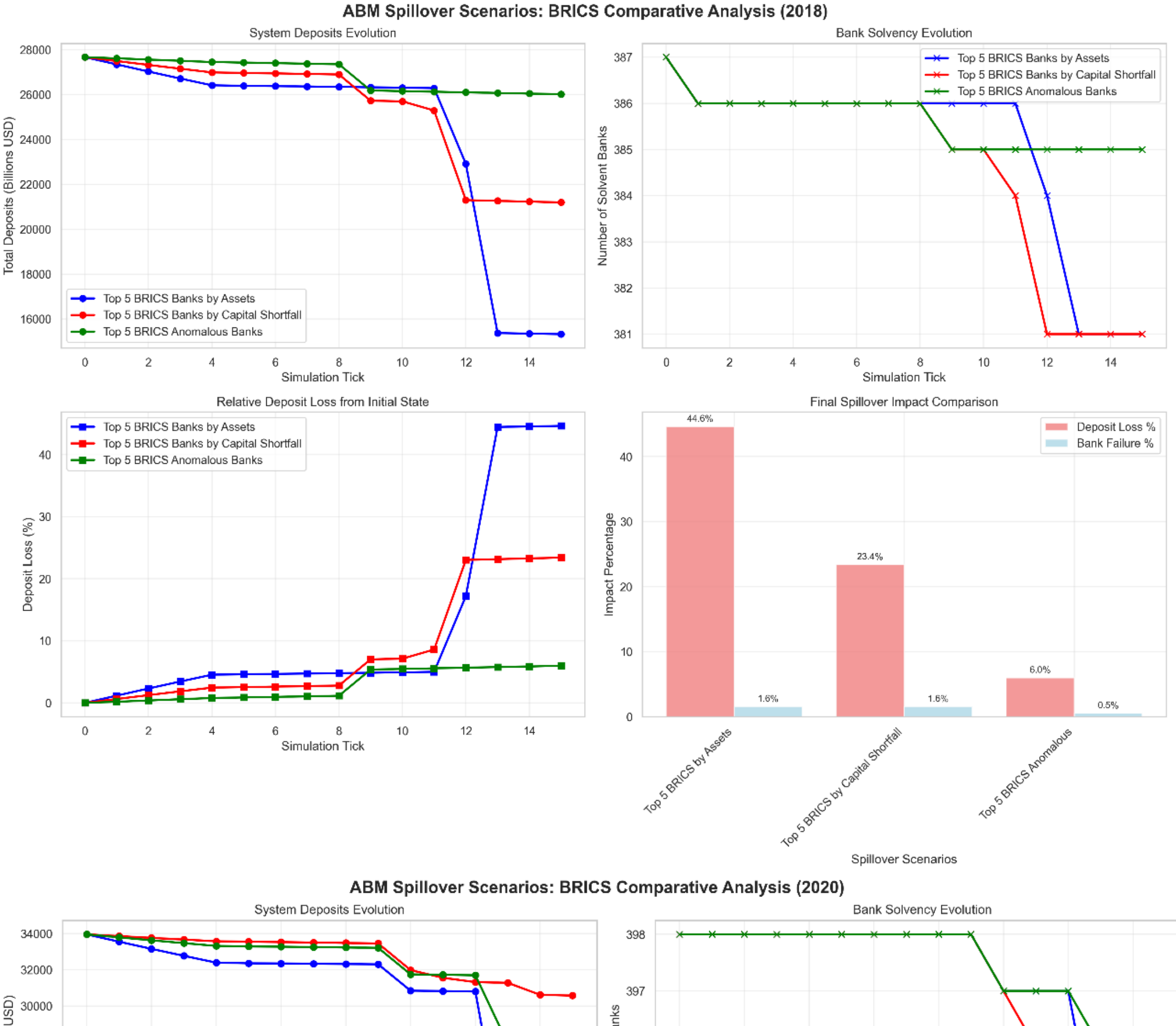

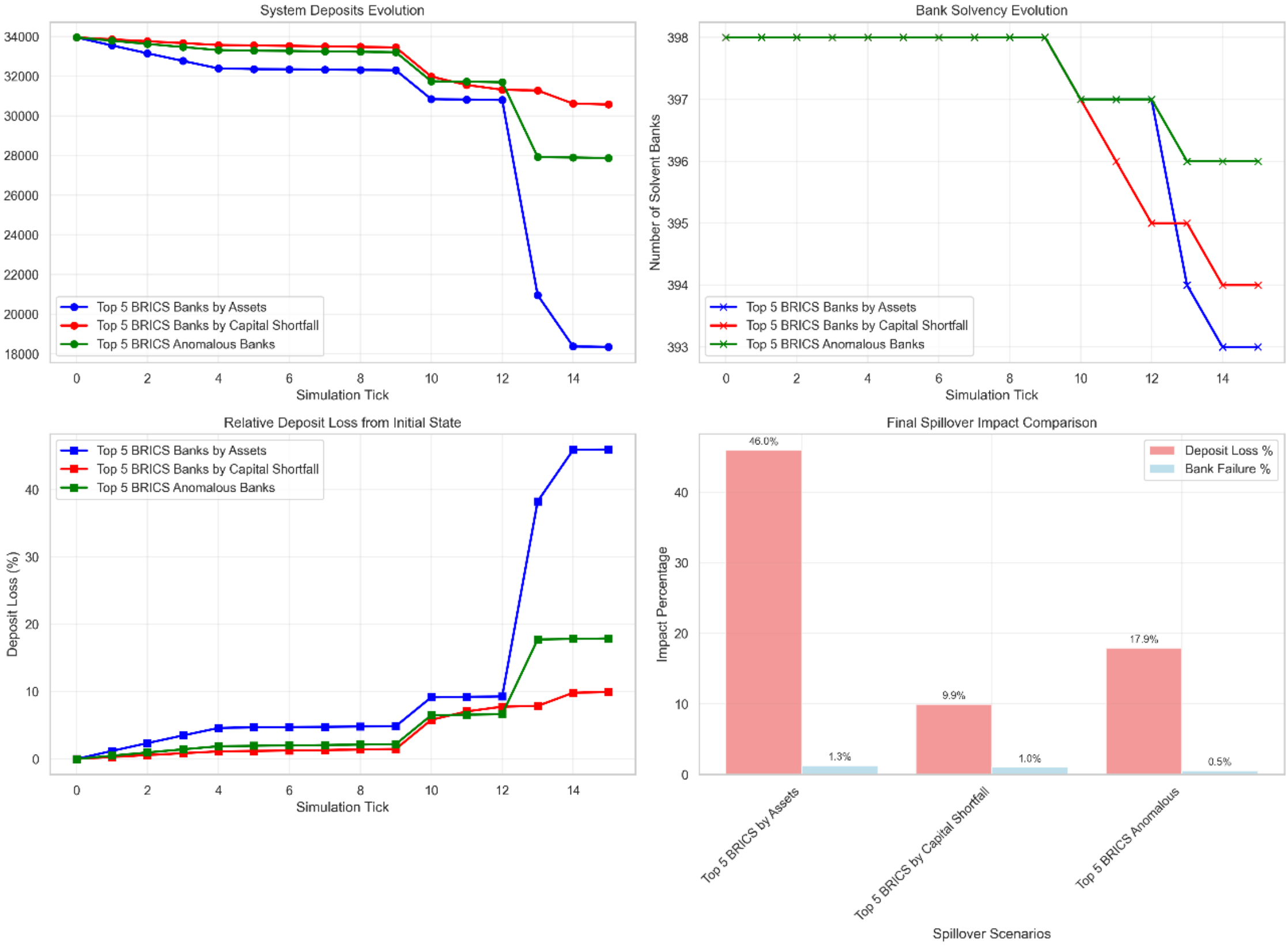

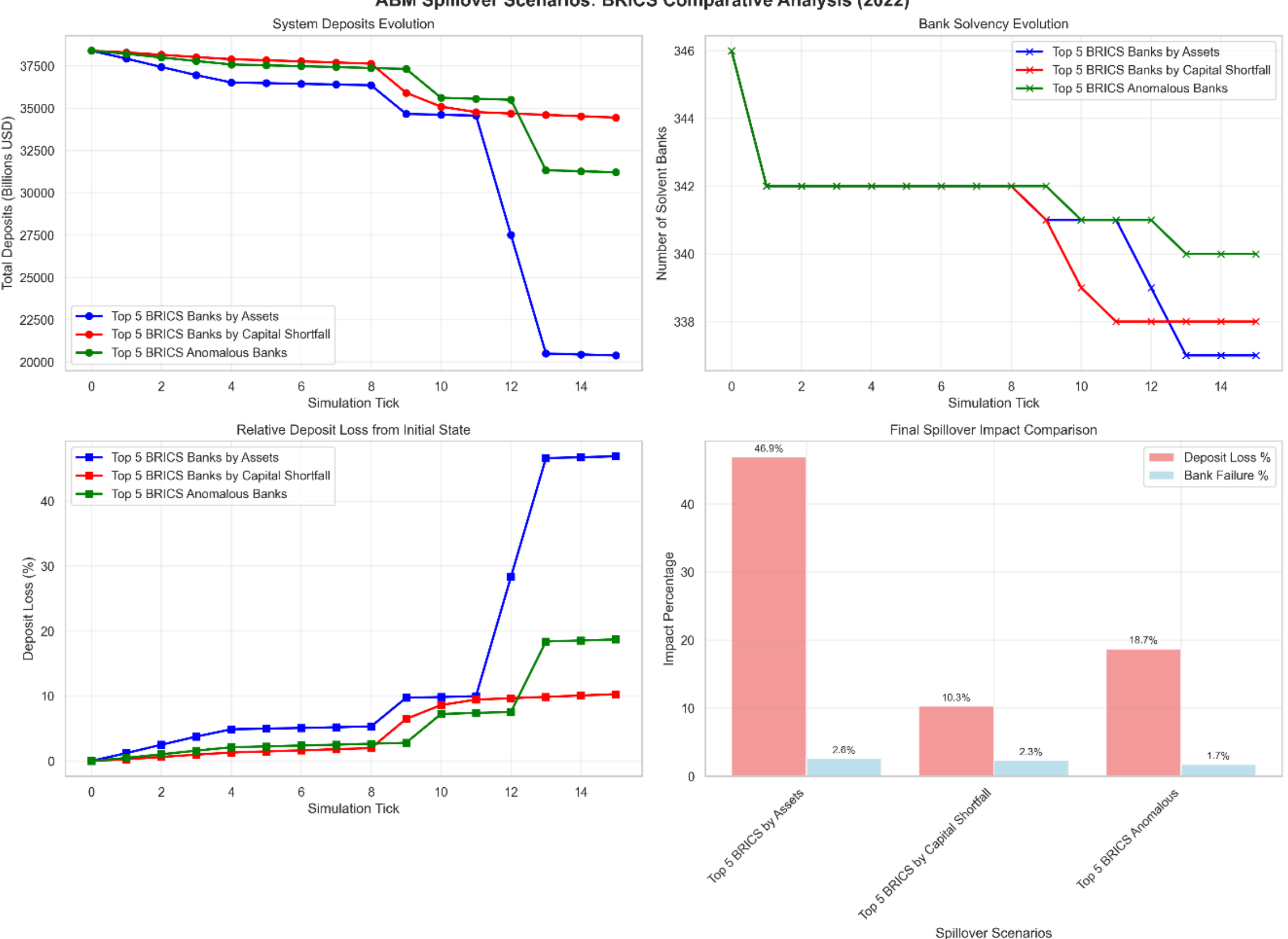

In each tested crisis year, the failure of the Top 5 Banks by Assets causes the greatest systemic damage. In the 2018 trade war simulation, for instance, shocking the largest banks resulted in a 44.6% system-wide deposit loss. We then compare this baseline to other risk profiles, finding that stressing the five most financially vulnerable institutions or the five most anomalous banks produced much lower network declines of 23.4% and 6.0%, respectively. Such a distinct gap in impact appears to support **Hypothesis H1**. The extent of systemic damage is not simply proportional to an individual bank's initial financial weakness. Instead, the sheer magnitude of the failure signal seems to drive the ensuing fallout. A G-SIBs bank's collapse acts as a coordinating signal for depositor panic. That initial fear may prompt a self-sustaining run across the entire network, suggesting behavioral contagion rather than direct structural exposure as the underlying mechanism.

In this study, we ran a final set of simulations to assess broader geopolitical threats, modeling scenarios where every bank in a specific country experiences simultaneous stress. The results shown in Figure 6 indicate that a correlated, national-level event causes more overall destruction than the combined losses of the five largest banks. These outcomes are likely to support **Hypothesis H2**. While the financial network might tolerate isolated institutional failures, it often fails to absorb broad external shocks. We test these systemic vulnerabilities against two historical situations: the Russian sanctions (encompassing both the 2014 and 2022 events) and the 2018 US-China trade war. The BRICS system faces a near-total collapse under such conditions. Simulated deposit losses reach nearly 100%, alongside bank failure rates exceeding 74%. Observing these patterns may suggest that the BRICS banking arrangement currently lacks the capacity to withstand coordinated stress targeting a member nation's entire financial sector.

**Figure 6**. ABM Geopolitical Shock Simulations: BRICS Countries Comparative Analysis

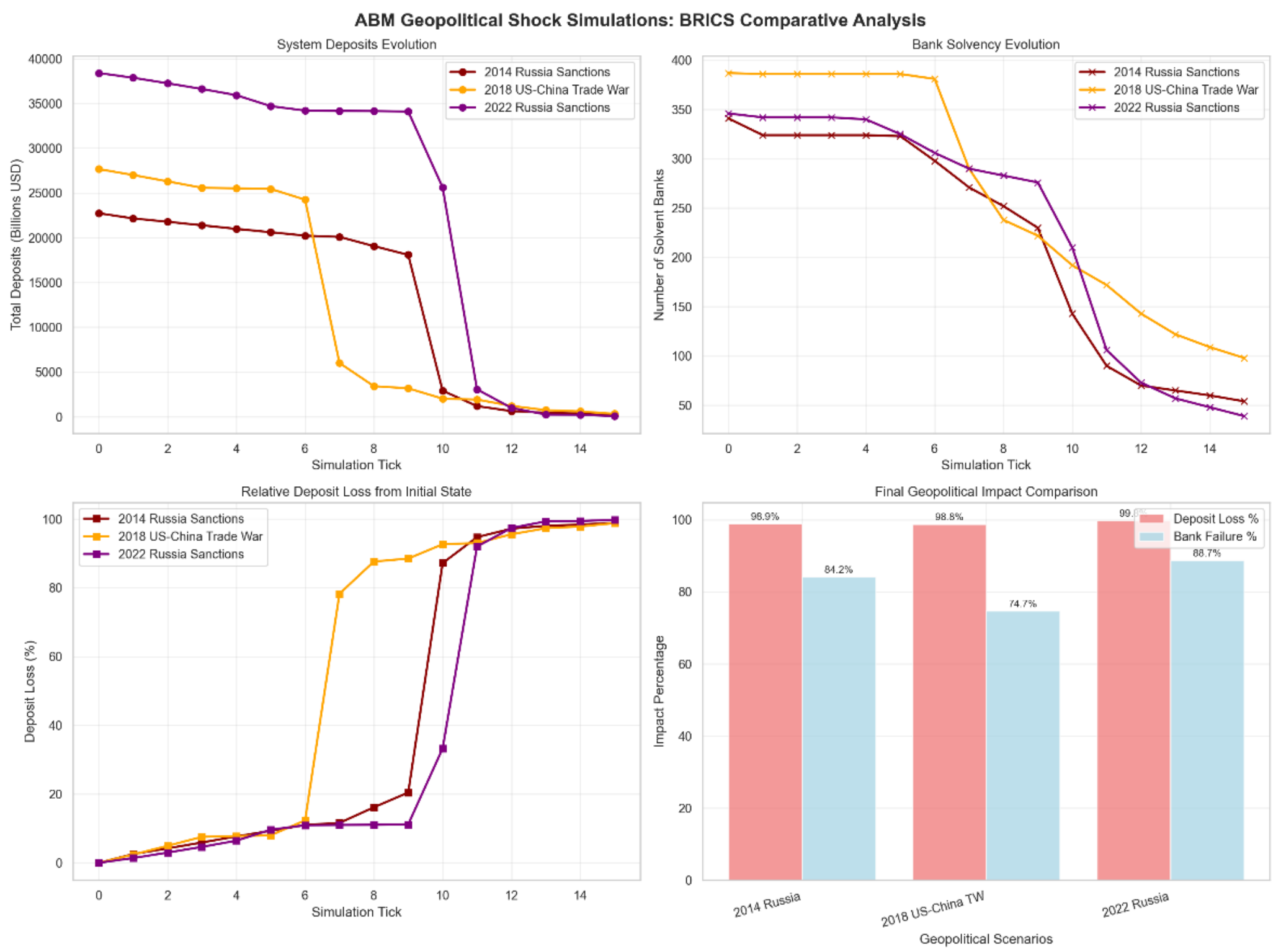

To address RQ2, we map annual changes in systemic risk using ABM simulations to track system resilience over time. Figure 7 visualizes these trends. China seems to act as a stable base for the group. During the Post-GFC period, India displayed a rapid recovery. Russia appears to be the least resilient member. The simulation results point to distinct structural vulnerabilities within the Russian system. Such a pattern contradicts the static SRISK_CS metric presented in Figure 5, which could mistakenly imply that Russia holds the lowest risk level.

We conduct annual Monte Carlo simulations of the ABM from 2018 to 2024 to evaluate the consistency of our resilience analysis. Table 6 and the online supplements detail the distribution of outcomes from 500 runs, which incorporate minor behavioral variations among agents. A relatively steady pattern emerges for most members. Brazil and China fall into the "Low Risk" classification. India also shares this stable profile, while South Africa tends to stay in the "Medium Risk" group. The data suggests a sharp drop in the Russian banking system's resilience over this same period. Between 2018 and 2020, Russia moved into the "Medium Risk" band. The country then shifted to "High Risk" by 2021 as its mean capital ratio declined to approximately 45%. A drop of this magnitude is likely to signal a major structural break in the local banking sector. The BRIDGES framework identified this severe system fragility an entire year before the 2022 invasion of Ukraine and the resulting global sanctions.

We should point out that the BRICS is far from a uniform entity like the Eurozone. The BRIDGES framework treats the BRICS as distinct macroeconomic and regulatory environments rather than a single standardized group, generating direct computational reflections of each country's underlying reality. For instance, China's high resilience is the result of its specific institutional arrangement, which relies on a closed capital account and the PBOC's capacity to quickly inject administrative liquidity into state-owned G-SIBs under tight state control. Brazil and South Africa operate market-driven banking systems that are deeply reliant on the global capital flows. Both nations have distinct contagion dynamics that depend heavily on external funding liquidity. The extreme fragility observed in Russia after 2021 likely stems from its increasing geopolitical isolation and the abrupt withdrawal of foreign capital.

**Figure 7: Systemic Resilience by BRICS countries (ABM Simulation Results, 2008-2024)**

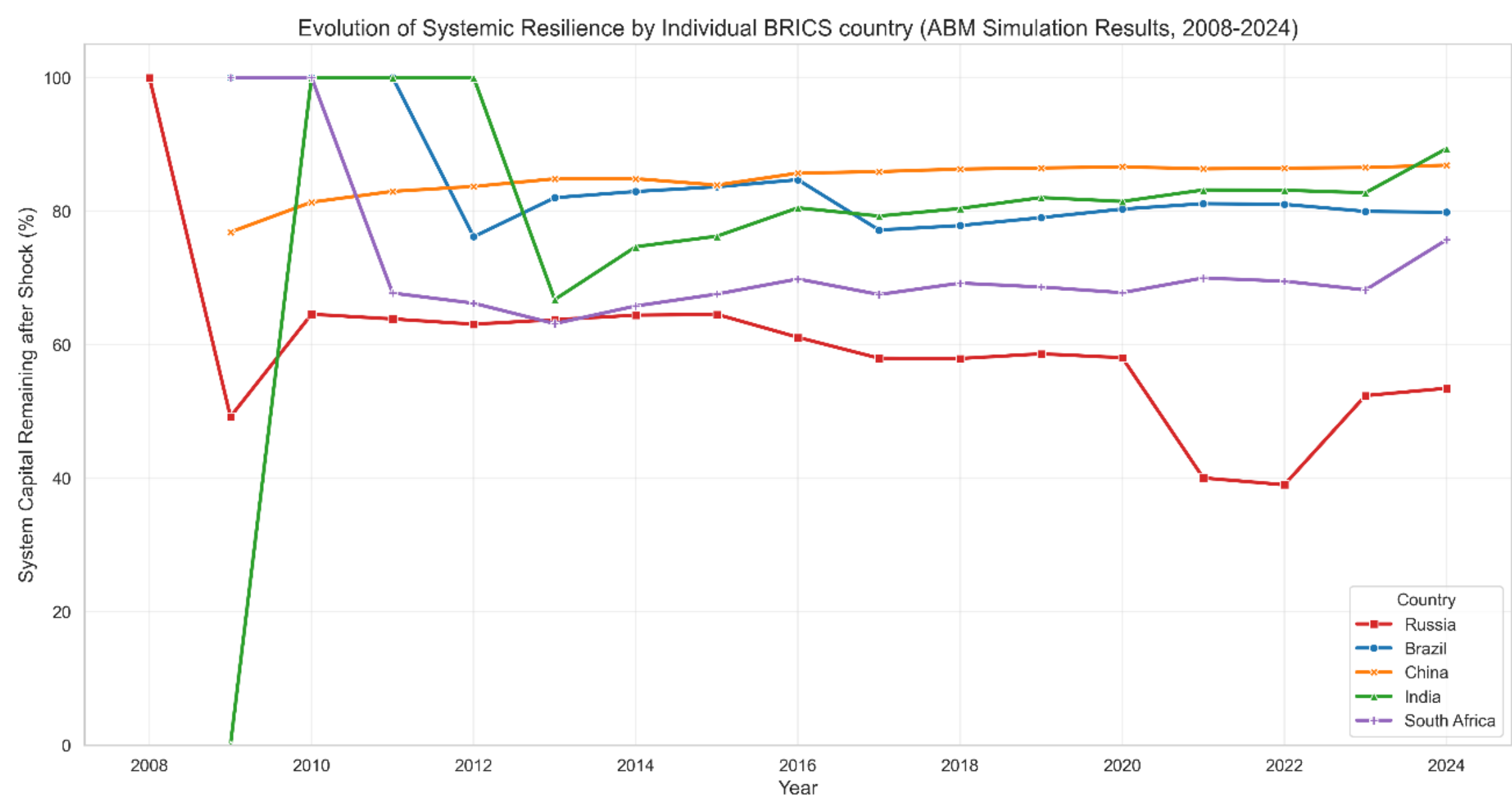

**Table 6. Probabilistic Country-Level Risk Classification (2018-2024)**

| Country | Risk Level | Year | | | | | | |
|---|---|---|---|---|---|---|---|---|
| | | 2018 | 2019 | 2020 | 2021 | 2022 | 2023 | 2024 |
| Brazil | Low | 0.24 | 0.22 | 1.00 | 1.00 | 1.00 | 0.22 | 0.27 |
| | Medium | 0.76 | 0.78 | 0.00 | 0.00 | 0.00 | 0.78 | 0.73 |
| China | Low | 1.00 | 1.00 | 1.00 | 1.00 | 1.00 | 1.00 | 1.00 |
| India | Low | 1.00 | 1.00 | 1.00 | 1.00 | 1.00 | 1.00 | 1.00 |
| | Low | 0.15 | 0.20 | 0.14 | 0.00 | 0.00 | 0.03 | 0.06 |
| Russia | Medium | 0.78 | 0.74 | 0.79 | 0.22 | 0.25 | 0.21 | 0.17 |
| | High | 0.07 | 0.06 | 0.07 | 0.78 | 0.00 | 0.76 | 0.77 |
| | Critical | 0.00 | 0.00 | 0.00 | 0.00 | 0.75 | 0.00 | 0.00 |
| South | Low | 0.22 | 0.24 | 0.24 | 0.21 | 0.25 | 0.22 | 0.23 |
| Africa | Medium | 0.78 | 0.76 | 0.76 | 0.79 | 0.75 | 0.78 | 0.77 |

### 4.3. Implications

One advantage of the BRIDGES framework is its ability to connect slow-moving structural shifts in the financial system to the probability of sudden shock events. The methodology integrates four distinct stages of analysis to build an evolving context. First, we establish a static baseline of vulnerability using metrics like SRISK_CS to flag potentially weak banks. Second, DTW network analysis maps the temporal connections among these institutions. Third, a TGNN evaluates the shifting network topology to detect periods of unusual activity. Fourth, an ABM simulation serves as a computational stress test. This final stage tests whether a basic first-order vulnerability is likely to trigger systemic failure under non-linear crisis conditions.

Such comparative analysis may represent the most useful contribution of our approach. Across the evaluated crisis years, the simulations generally indicate that shocking the largest institutions causes greater systemic damage than targeting the most vulnerable or anomalous banks. This outcome suggests that the "too big to fail" problem involves behavioral factors alongside basic financial architecture. A panic effect likely drives these widespread collapses. A second-order behavioral cascade seems to outweigh the expected failure of strictly undercapitalized banks.

Regulators might find the BRIDGES framework useful beyond basic capital adequacy measures such as SRISK. The approach allows for active monitoring of shifting network centrality, which appears helpful for anticipating rapid, behavior-driven panics. Traditional accounting-based methods, which typically rely on annual balance sheet data, fall short when calculating the time-varying metrics required for SRISK. The BRIDGES framework addresses this shortcoming by incorporating SRISK_CS within a dynamic simulation environment. The framework can assess when and how this static vulnerability translates into systemic damage by subjecting banks with high SRISK_CS to various geopolitical shocks.

DTW distance seems to handle the irregularities of financial reporting better than cosine similarity or Euclidean distance. It is computationally expensive, though. Increasing the number of banks and observations appears to increase resource requirements rapidly. For high-frequency market data, cosine similarity and Euclidean metrics might be preferred when computing budgets are tight. In this study, we aligned our sample size with available computing capacity, and running the BRIDGES framework took less than 120 minutes.

## 5. Conclusion

In this study, we introduced the BRIDGES framework to evaluate long-term structural risk within the BRICS banking systems. The results point to four main observations. First, when modeling endogenous financial failures, the behavioral influence of a bank's size appears to be the primary driver of systemic collapse. Second, simulations suggest that the panic caused by the collapse of a "too big to fail" bank is far more destructive than that from the failure of a financially weaker or anomalous bank. Third, a large-scale, correlated geopolitical shock targeting a central member poses severe tail risk to the system. Fourth, the model indicates that such a geopolitical event is likely to trigger a near-total collapse, creating damage that easily exceeds any isolated banking failure.

These findings may suggest specific policy adjustments. For Chinese authorities, strictly regulating mega-banks based on capital adequacy might be insufficient on its own. Regulators appear to need a stronger focus on managing the psychological and signaling effects of these banks (G-SIBs), given that their visible distress is likely to cause large behavioral panics across the BRICS banking systems.

We rely exclusively on annual balance sheet data, which presents a clear limitation. Including unlisted state-owned banks, this choice was made necessary. That said, the approach sacrifices the fast responsiveness often found in market-based frameworks. Future extensions of the BRIDGES framework might adopt a mixed method to connect structural accounting records with immediate market sentiment. First, high-frequency market prices could help calibrate ABM's fear dynamics for publicly listed banks. Second, the DTW-balance sheet method is likely to remain the standard for unlisted banks. This approach is likely to create a hybrid model that connects immediate market sentiment with actual balance-sheet conditions. Also, endogenizing the behavioral parameters $(\alpha, \psi)$ over time, based on current social media sentiment or macroeconomic volatility indices, could be valuable. Doing so appears to offer an even more exact, immediate warning system for regulators in emerging markets.

## References

Acharya, V., R. Engle, M. Richardson, 2012, Capital Shortfall: A New Approach to Ranking and Regulating Systemic Risks. American Economic Review 102, 59-64.

Ahmad, A., A. Khan, S. Akhtar, H.W. Akram, 2021, Examining the Development of Banking Sector Regulations and Supervision Practices across BRICS and G7 Countries. Complexity 2021, 1192829.

Ahmad, W., A.V. Mishra, K.J. Daly, 2018, Financial connectedness of BRICS and global sovereign bond markets. Emerging Markets Review 37, 1-16.

Allen, F., A. Babus, E. Carletti, 2012, Asset commonality, debt maturity and systemic risk. Journal of Financial Economics 104, 519-534.

An, H., H. Wang, S. Delpachitra, S. Cottrell, X. Yu, 2022, Early warning system for risk of external liquidity shock in BRICS countries. Emerging Markets Review 51, 100878.

Armijo, L.E., C. Roberts, 2014. The emerging powers and global governance: Why the BRICS matter, Handbook of emerging economies. (Routledge), pp. 503-524.

Barik, R., A.K. Pradhan, 2021, Does financial inclusion affect financial stability: evidence from BRICS nations? The journal of developing areas 55.
Berndt, D.J., J. Clifford, 1994. Using dynamic time warping to find patterns in time series, Proceedings of the 3rd International Conference on Knowledge Discovery and Data Mining. (AAAI Press, Seattle, WA), pp. 359–370.
Bookstaber, R., 2017, Agent-Based Models for Financial Crises. Annual Review of Financial Economics 9, 85-100.
Bouvatier, V., A. López-Villavicencio, V. Mignon, 2012, Does the banking sector structure matter for credit procyclicality? Economic Modelling 29, 1035-1044.
Boyd, J.H., G. De NicolÓ, 2005, The Theory of Bank Risk Taking and Competition Revisited. The Journal of Finance 60, 1329-1343.
Brownlees, C., R.F. Engle, 2017, SRISK: A Conditional Capital Shortfall Measure of Systemic Risk. The Review of Financial Studies 30, 48-79.
Cai, J., F. Eidam, A. Saunders, S. Steffen, 2018, Syndication, interconnectedness, and systemic risk. Journal of Financial Stability 34, 105-120.
Cerutti, E., C. Casanova, S.-K. Pradhan, 2018, The growing footprint of EME banks in the international banking system. BIS Quarterly Review December.
Chu, Y., S. Deng, C. Xia, 2020, Bank Geographic Diversification and Systemic Risk. The Review of Financial Studies 33, 4811-4838.
Diamond, D.W., P.H. Dybvig, 1983, Bank Runs, Deposit Insurance, and Liquidity. Journal of Political Economy 91, 401-419.
Engle, R., 2018, Systemic Risk 10 Years Later. Annual Review of Financial Economics 10, 125-152.
Engle, R., C. Zazzara, 2018, Systemic risk in the financial system: Capital shortfalls under Brexit, the US elections and the Italian referendum. Journal of Credit Risk 14, 97-120.
FSB, 2024, 2024 List of Global Systemically Important Banks (G-SIBs), https://www.fsb.org/uploads/P261124.pdf, Accessed on October 12, 2025
Girardi, G., K.W. Hanley, S. Nikolova, L. Pelizzon, M.G. Sherman, 2021, Portfolio similarity and asset liquidation in the insurance industry. Journal of Financial Economics 142, 69-96.
Goldstein, I., A.D.Y. Pauzner, 2005, Demand–Deposit Contracts and the Probability of Bank Runs. The Journal of Finance 60, 1293-1327.
Horn, S., C.M. Reinhart, C. Trebesch, 2021, China's overseas lending. Journal of International Economics 133, 103539.
Huang, X., H. Zhou, H. Zhu, 2012, Systemic risk contributions. Journal of Financial Services Research 42, 55-83.
Kukacka, J., L. Kristoufek, 2020, Do ‘complex’ financial models really lead to complex dynamics? Agent-based models and multifractality. Journal of Economic Dynamics and Control 113, 103855.
Lall, R., 2009. Why Basel II failed and why any Basel III is doomed. (GEG working paper).
Lawson, C., 1994, THE THEORY OF STATE-OWNED ENTERPRISES IN MARKET ECONOMIES. Journal of Economic Surveys 8, 283-309.
Li, S., 2019, The Impact of Bank Regulation and Supervision on Competition: Evidence from Emerging Economies. Emerging Markets Finance and Trade 55, 2334-2364.
Lux, T., 2018, Estimation of agent-based models using sequential Monte Carlo methods. Journal of Economic Dynamics and Control 91, 391-408.

Marcus, A.J., 1984, Deregulation and bank financial policy. Journal of Banking & Finance 8, 557-565.
Mengistae, T., L. Colin Xu, 2004, Agency Theory and Executive Compensation: The Case of Chinese State-Owned Enterprises. Journal of Labor Economics 22, 615-637.
Mensi, W., S. Hammoudeh, D.K. Nguyen, S.H. Kang, 2016, Global financial crisis and spillover effects among the U.S. and BRICS stock markets. International Review of Economics & Finance 42, 257-276.
Minsky, H.P., 1970. Financial instability revisited: The economics of disaster. (Board of Governors of the Federal Reserve System St. Louis).
Moudud-Ul-Huq, S., 2020, Does bank competition matter for performance and risk-taking? empirical evidence from BRICS countries. International Journal of Emerging Markets 16, 409-447.
Orlik, T., T. Orlik, 2020, China: The bubble that never pops. (Oxford University Press).
Pareja, A., G. Domeniconi, J. Chen, T. Ma, T. Suzumura, H. Kanezashi, T. Kaler, T. Schardl, C. Leiserson, 2020. EvolveGCN: Evolving Graph Convolutional Networks for Dynamic Graphs, Proceedings of the AAAI Conference on Artificial Intelligence.), pp. 5363-5370.
Qin, X., C. Zhou, 2019, Financial structure and determinants of systemic risk contribution. Pacific-Basin Finance Journal 57, 101083.
Ramamurti, R., 1987, Performance Evaluation of State-Owned Enterprises in Theory and Practice. Management Science 33, 876-893.
Richards, T.J., R.N. Acharya, A. Kagan, 2008, Spatial competition and market power in banking. Journal of Economics and Business 60, 436-454.
Saliba, C., P. Farmanesh, S.A. Athari, 2023, Does country risk impact the banking sectors' non-performing loans? Evidence from BRICS emerging economies. Financial Innovation 9, 86.
Sehrawat, M., A.K. Giri, 2015, Financial development and economic growth: empirical evidence from India. Studies in Economics and Finance 32, 340-356.
Sharma, S., A. Anand, 2018, Income diversification and bank performance: evidence from BRICS nations. International Journal of Productivity and Performance Management 67, 1625-1639.
Singh, D., M. Theivanayaki, M. Ganeshwari, 2021, Examining Volatility Spillover Between Foreign Exchange Markets and Stock Markets of Countries such as BRICS Countries. Global Business Review 25, 1269-1289.
Siva Kiran Guptha, K., R. Prabhakar Rao, 2018, The causal relationship between financial development and economic growth: an experience with BRICS economies. Journal of Social and Economic Development 20, 308-326.
Sorkin, A.R., 2010, Too big to fail: The inside story of how Wall Street and Washington fought to save the financial system--and themselves. (Penguin).
Stern, G.H., R.J. Feldman, 2004, Too big to fail: The hazards of bank bailouts. (Rowman & Littlefield).
Strobl, G., 2022, A theory of procyclical market liquidity. Journal of Economic Dynamics and Control 138, 104326.
Stuenkel, O., 2020, The BRICS and the future of global order. (Bloomsbury Publishing PLC).
Umar, M., G. Sun, 2016, Interaction among funding liquidity, liquidity creation and stock liquidity of banks: Evidence from BRICS countries. Journal of Financial Regulation and Compliance 24, 430-452.

Umar, M., G. Sun, K. Shahzad, Z.-u.-R. Rao, 2018, Bank regulatory capital and liquidity creation: evidence from BRICS countries. International Journal of Emerging Markets 13, 218-230.
Vandin, A., D. Giachini, F. Lamperti, F. Chiaromonte, 2022, Automated and distributed statistical analysis of economic agent-based models. Journal of Economic Dynamics and Control 143, 104458.
Zeb, S., A. Rashid, 2019, Systemic risk in financial institutions of BRICS: measurement and identification of firm-specific determinants. Risk Management 21, 243-264.